\documentclass[11]{article}
\usepackage[dvips]{graphicx}
\usepackage{authblk}
\usepackage{dcolumn}
\usepackage{bm}
\usepackage{cite}
\usepackage{color}
\usepackage{soul}
\usepackage{hyperref}
\usepackage{amsmath}
\usepackage{geometry}
\usepackage{indentfirst}
\usepackage[english]{babel}
\usepackage[utf8]{inputenc}
\usepackage{fancyhdr}

\begin{document}
\title{A building block for hardware belief networks}
\author{Behtash Behin-Aein}
\affil[1]{GLOBALFOUNDRIES Inc. USA, Santa Clara, CA 95054}
\author[2]{Vinh Diep}
\affil[2]{School of ECE, Purdue University, West Lafayette, IN 47907}
\author[2]{Supriyo Datta}
\maketitle
\providecommand{\keywords}[1]{\textbf{\textit{Index terms---}} #1}

\begin{abstract}
Belief networks represent a powerful approach to problems involving probabilistic inference, but much of the work in this area is software based utilizing standard deterministic hardware based on the transistor which provides the gain and directionality needed to interconnect billions of them into useful networks. This paper proposes a transistor like device that could provide an analogous building block for probabilistic networks. We present two proof-of-concept examples of belief networks, one reciprocal and one non-reciprocal, implemented using the proposed device which is simulated using experimentally benchmarked models.
\end{abstract}
\keywords{stochastic, sigmoid, phase transition, spin glass, frustration, reduced frustration, Ising model, Bayesian network, Boltzmann machine}

\tableofcontents
\pagebreak
\section{Introduction}
\begin{figure*}[]
  \centering
   \includegraphics[width=7.5cm, height=7cm]{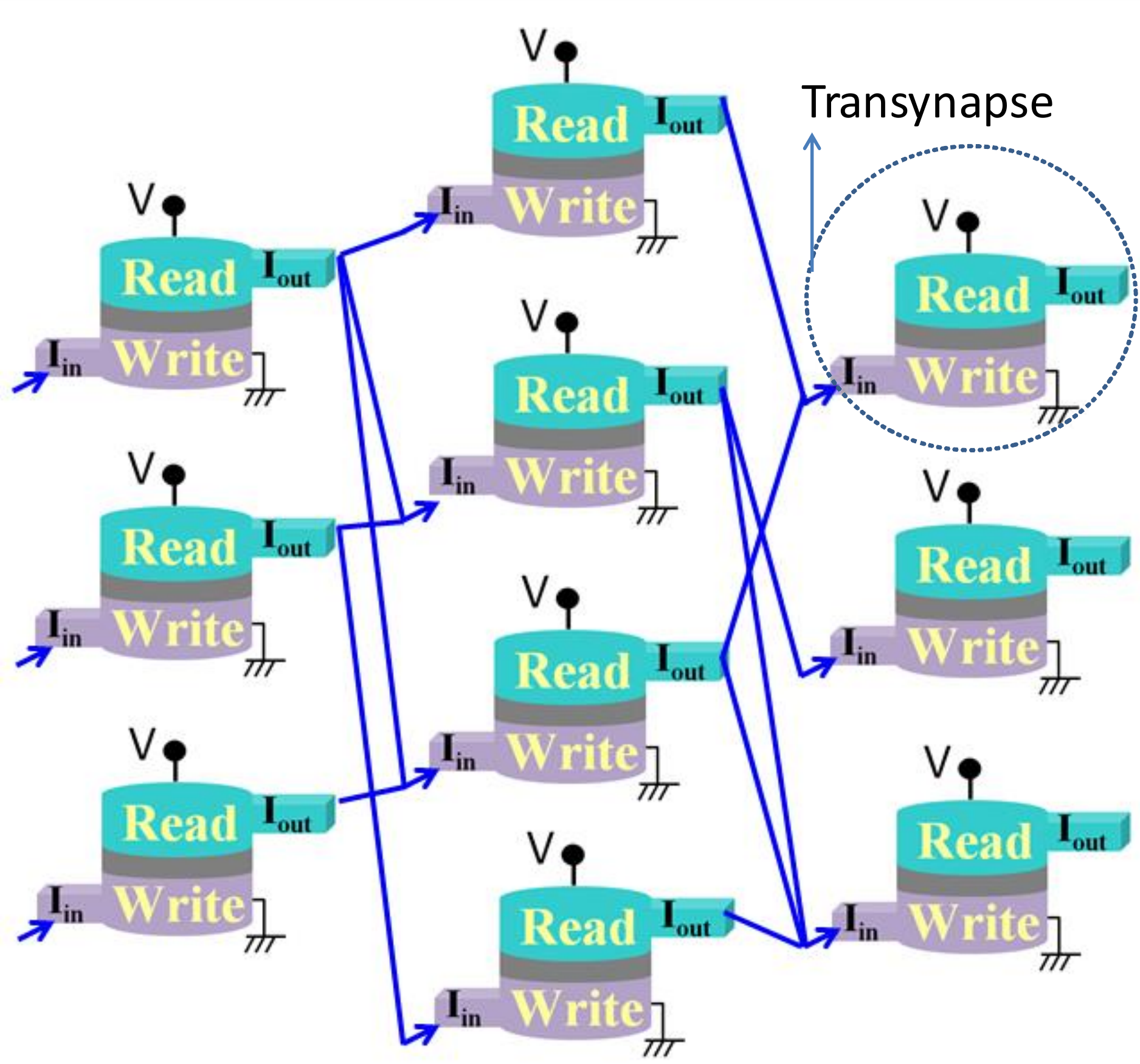}\\
  \caption{This paper defines a transynapse that can be interconnected to build probabilistic networks as shown schematically.
  Section 2 describes a specific transynapse design based on experimentally benchmarked models which are then used to illustrate
  the use of transynapse networks to solve problems involving belief networks in sections 3 and 4.}
  \label{F1}
\end{figure*}

\paragraph{} Probabilistic computing is a thriving field of computer science and mathematics and is widely viewed as a powerful approach for tackling the daunting problems of searching, detection and inference posed by the ever increasing amount of ``big data''. \cite{Hinton2006NC, Hinton2006Sc, BengioGreedy, BengioAI, PearlBook, PearlCausality, BengioReview, Hinton2002PoE,CDBN,Smolensky,Hinton1985} Much of this work, however, is software-based, utilizing standard general purpose hardware that is based on high precision deterministic logic.\cite{HNN} The building block for this standard hardware is the ubiquitous transistor which has the key properties of gain and directionality  that allow billions of them to be interconnected to perform complex tasks. This paper proposes a transistor-like device that could provide an analogous building block for probabilistic logic.
\paragraph{} A number of authors \cite{KR,CK,SMS} have recognized that the physics of nanomagnets can be exploited for stochastic logic and natural random number generators to replace the complex circuitry that is normally used. However, these are individual stochastic circuit elements within the standard framework of complementary metal oxide semiconductor (CMOS) transistors, which provide the necessary gain and directionality. By contrast, what we are proposing in this paper are networks constructed out of magnet-based stochastic devices that have been individually engineered to provide transistor-like gain and directionality so that they can be used to construct large scale circuits without external transistors (Fig.\ref{F1}a).
\paragraph{} Feynman (1982) alluded to a probabilistic computer based on probabilistic hardware that could efficiently solve problems
involving classical probability, contrasting it with a quantum computer based on quantum hardware that could efficiently solve quantum
problems. This paper inspired much work on quantum computing, but we would like to draw attention to his description of a probabilistic
computer:
\linebreak``\textit{\ldots the other way to simulate a probabilistic nature, which I'll call N .. is by a computer C which itself is
probabilistic, .. in which the output is not a unique function of the input. \ldots it simulates nature in this sense: that C goes from some
.. initial state .. to some final state with the same probability that N goes from the corresponding initial state to the corresponding final
state. \ldots If you repeat the same experiment in the computer a large number of times \ldots it will give the frequency of a given final
state proportional to the number of times, with approximately the same rate \ldots as it happens in nature.}'' The possibility of
probabilistic computing machines has also been addressed by more recent authors. \cite{SB,FBM,BBFrance,BBPatent, Murray1, Murray2} The
primary purpose of this paper is to introduce the concept of a `transynapse', a device that can be interconnected in large numbers to build
probabilistic computers (Fig.\ref{F1}).  \linebreak
\paragraph{} The transynapse combines a synapse-like function with a transistor-like gain and directionality and in Section 2 we describe a device that uses the established physics of nanomagnets to implement it. We present a specific design for the transynapse which is simulated using experimentally benchmarked models (supplementary section $1$) for established phenomena to demonstrate the stochastic sigmoid transfer function. In section 3 we use these same models to show how transynapses can be used to build either of two fundamentally different class of networks, an Ising like network with symmetric network interactions and a non-Ising network with directed interactions. In Section 4 we then present two proof-of-concept examples of belief networks, one reciprocal and the other non-reciprocal, implemented using transynapses which are simulated with the same experimentally validated models used in Section 2.
\pagebreak
\section{Transynapse: The Building Block}
\begin{figure*}[]
  \centering
   \includegraphics[width=12cm, height=6cm]{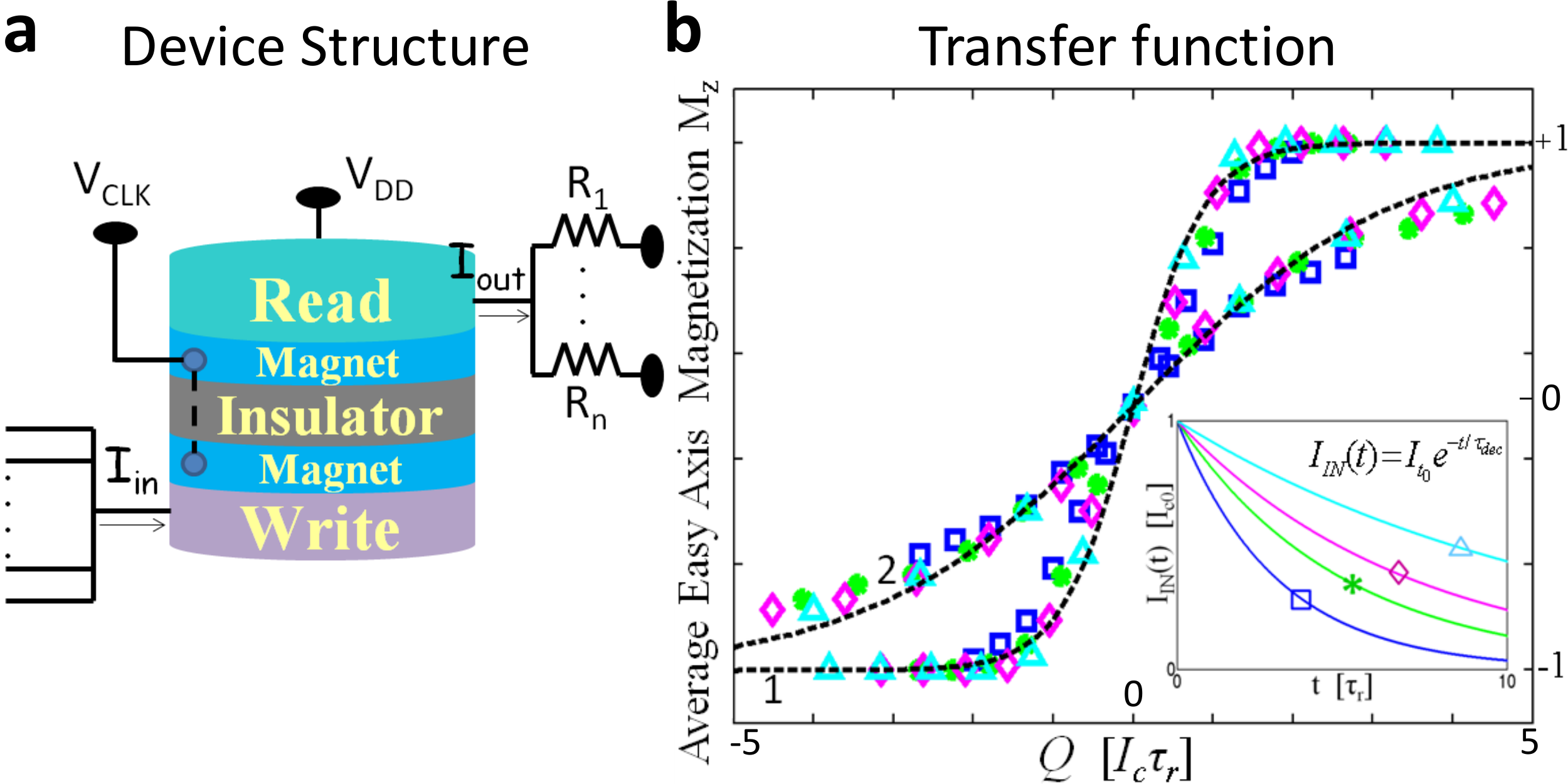}\\
  \caption{Design for a transynapse: (a)Device structure: For our simulations we use the same design as that in Datta et al.
  \cite{DattaSS} which provides the required gain, fan-in and fan-out, making use of the established physics of the spin Hall
  effect (SHE) for the input and the magnetic tunnel junction (MTJ) for the output. (see also \cite{DattaChap},\cite{BehtashMRS})
  However, instead of the deterministic mode described earlier, we operate it in a probabilistic mode as described next.
  \textbf{(b)} The WRITE and READ magnets are both initialized along their hard axis and allowed to relax in the presence of an
  exponentially decaying current (see inset) $I_{IN}(t)$ and decay time parameter $\tau_{dec}$. The outputs obtained from a
  statistical average of 1000 Monte Carlo runs for different input currents all fall on a single universal curve when plotted
  against $Q$, the time integrated current weighted by the factor $e^{-t/\tau_r}$. Curves 1 and 2 are obtained for nanomagnets with
  energy barriers $48k_BT$ and $12k_BT$ respectively and are described well by Eq.\ref{M_z}.}
  \label{F2}
\end{figure*}

The transynapse consists of a WRITE unit and a READ unit (Figs.\ref{F1},\ref{F2}), electrically isolated from each other. The WRITE unit of transynapse $T_i$ sums a set of input signals $I_{IN,j}(t)$, integrates them over a characteristic time scale $\tau_r$ to obtain a quantity
\begin{subequations}\label{Iin&Iout}
   \begin{align}
	Q_i&=\int_{0}^{\infty}dtI_{IN,i}(t)e^{-t/\tau_r}    \label{Q} \\
    \intertext{which determines the \textit{mean value} of state $S$ of the device through a sigmoidal function of the form}
	\bar{S}(Q_i)&= \tanh \left(\frac{Q_i}{2I_0\tau_r}\right)        \label{S} \\
	\intertext{where $I_0$ is a characteristic current. The READ unit produces multiple weighted output currents proportional to the average state $\bar{S}$:}
    \bar{I}_{OUT,j}(Q)&= I_0 w_{ji} \bar{S}(Q_i)             \label{Iout,j}
    \end{align}
\end{subequations}

\paragraph{} In this paper we use a specific design for this device following that described in Datta et al.\cite{DattaSS} which uses an input WRITE unit magnetically coupled to, but electrically isolated from an output READ unit. It provides the required gain, fan-in and fan-out, making use of experimentally benchmarked models (supplementary section $1$) for the established physics of the spin Hall effect (SHE) for the input and the magnetic tunnel junction (MTJ) for the output. However, for transynapse operation, we need to operate it in a probabilistic mode not considered before, where the input and output are not deterministic variables but stochastic ones. This could be done by using nanomagnets with low energy barriers ($E_b < 5k_BT$) that are in the super paramagnetic regime considering long enough programming times. The output (y-axis in Fig.2b) could then be interpreted as the time averaged magnetization of a single magnet along its easy axis. However, in this paper we use a different approach as explained below.
\paragraph{} For the simulations presented in this paper, nano-magnets are initialized along their hard
axis at t=0 and then allowed to relax. The output is obtained from a statistical average of the
magnetization $M_z$ along the easy axis obtained from 1000 Monte Carlo runs based on the stochastic Landau-Lifshitz-Gilbert (LLG)
equation, one for each magnet (WRITE and READ) coupled through a magnetic interaction as in
\cite{DattaSS}. The details are described in supplementary section $1$. Fig.\ref{F2}b shows that the
numerically obtained average $M_z$ is described well by the relation
\begin{subequations}
   \begin{align}
   \bar{M_z}(Q) &= \tanh \left(\frac{Q}{2I_0\tau_r}\right) \label{M_z}      \\
   \nonumber
   \end{align}
\end{subequations}
where $\tau_r = f_T(1+\alpha^2) / (2\alpha \gamma H_k)$, $H_K$ being the anisotropy field, $\gamma$ , the gyromagnetic ratio and $\alpha$ is the damping parameter. Also $I_0=\frac{I_c}{\eta}$ , where $I_c$ is the switching current for the nanomagnet\cite{Sun}, while the factor $\eta$ depends on its energy barrier $E_b$ and is given by the relation $\eta\approx0.06(E_b/k_BT)^{0.94}$ obtained from numerical simulations. Factor $f_T$ (6 for $E_b=48k_BT$ and 24 for $E_b=12k_BT$) determines how fast the magnetization relaxes depending on the ambient temperature. The results were obtained using different (Fig.\ref{F2}b inset) input currents of the form $I_{IN}(t)=I_{t0}e^{-t/\tau_{dec}}$, but the resulting output is well described by a single curve $M_z(Q)$ irrespective of the amplitude $I_{t0}$ and decay time parameter $\tau_{dec}$. This independence to time-decay parameters suggests that the probability of finalizing a magnet in one of the two
states essentially depends on the number ($N_s$ where $Q \propto N_s$) of Bohr magnetons (units of electron spin) imparted on it.
\cite{BehtashAPL} Indeed, similar underlying principles have been demonstrated experimentally in Ref.\cite{Kent_DtoT} in somewhat
different set ups for switching magnets from one state to the other in the short pulse regimes well above $I_c$. \par
\paragraph{} It is important to note the key attributes of the device that are needed to enable the construction of belief networks
by interconnecting hundreds of devices. Firstly, it is important to ensure input-output isolation, which is achieved by having
magnetically coupled WRITE and READ magnets separated by an insulator as shown. This separation would not be needed if the magnet
itself were insulating (like YIG, Yttrium iron garnet). The second important attribute is its gain defined as the maximum output charge current relative
to the minimum charge current needed to swing the probability from 0.5 (fully stochastic) to 1 (fully deterministic). This is the
quantity that determines the maximum fan-out that is possible which is particularly important if we want a high degree of inter
connectivity. The physics of SHE \cite{Ralph, SayeefGSHE} helps provide gain since for each device, it can be designed \cite{DattaSS,DattaChap,BehtashMRS} to provide more spin current to the WRITE magnet than the charge current provided by the READ unit of the preceding stage. \par
\paragraph{} The third attribute of the proposed device is its ability to sum multiple inputs and this can be done conveniently
since it is current-driven. A WRITE circuit consisting of a SHE metal like Tantalum provides a current-driven low impedance input,
different from the voltage-driven high input impedance field-effect transistors (FET's). The low input impedance ensures that the
total current into the WRITE unit is determined by the output impedance of the READ units of preceding stages
\cite{DattaSS,DattaChap,BehtashMRS}. This impedance is set by the intrinsic resistance of the READ units which could be on the
order of a $k\Omega$  if using magnetic tunnel junctions (MTJ's) or could be much lower if using the inverse spin Hall effect
(ISHE).\cite{ISHE} In either case an external series resistor R could be used to raise the output impedance as shown in Fig.
\ref{F2}a.
\begin{equation}
   \hspace{2cm} \bar{I}_{OUT,j}(Q)= \frac{V_{DD}(R_{AP}-R_{P})}{(R_{AP}+R_P)(R_{AP}+R_P+R_j+R_{IN})}\bar{M_z}(Q) \hspace{2.0cm}
   \text{(2b)} \label{Iout}     \nonumber
\end{equation}
$V_{DD}$ being the external voltage, $R_P$ and $R_{AP}$, the parallel and anti-parallel resistance of the MTJ, $R_{IN}$, the input
resistance of the next device and $R_j$ is the external series resistance which can be used to weight the outputs appropriately.
The weighting of the output can also be accomplished by tuning $V_{DD}$ where multiple bipolar output weights sharing the same input
can be implemented via a common WRITE unit with multiple READ units as shown in Fig.\ref{F5}a. \cite{Vinh} \par
\paragraph{}We envision that the detailed physics used to implement the transynapse will evolve, especially the physics used for
the WRITE, the READ and/or the weighting, since this field is in a stage of rapid development with new discoveries being reported
on a regular basis. The input (or WRITE) circuit could utilize phenomena other than the SHE used here, just as the output (or READ)
circuit could use mechanisms other than MTJ’s. Similarly, the nanomagnet can be initialized in a neutral state with modern voltage
driven mechanisms \cite{Chien, Pedram} like voltage controlled anisotropy, or with established methods like an external magnetic
field \cite{Imre,BehtashSE} or spin torque \cite{Brataas,ASL,SayeefGSHE}, or thermal assistance \cite{TAS}. Alternatively, as
mentioned earlier nanomagnets in the super paramagnetic regime could be used with the mean state $\bar{S}$ defined by a time
average instead of an ensemble average. The purpose of this manuscript is simply to establish the general concept of a transynapse
that integrates a synapse-like behavior with a transistor-like gain and isolation, thus permitting the construction of compact
large scale belief networks.\par
\paragraph{}Note also that our transynapses are assumed to communicate via charge current since that is a well-established robust form of communication. However, communication could be influenced through spin channels (as in all-spin logic \cite{ASL,BBFrance,BBPatent}) or through spin waves requiring very different WRITE and READ units. \par
\pagebreak
\section{Reciprocal and \textit{Non-Reciprocal} Networks}
\paragraph{}  A key feature of transynapse is the flexibility it affords in adjusting the weight $w_{ji}$ that determines the influence of one transynapse ($T_i$) on another ($T_j$), by adjusting the parameters of the READ unit of $T_i$. The weight $w_{ij}$ on the other hand is controlled independently through the READ unit of $T_j$. If we choose $w_{ij}=w_{ji}$, we have a bidirectional or reciprocal network similar to the type described by an Ising model described by a Hamiltonian $H$. In such networks the probability $P_n$ of a specific configuration, $n=\left\lbrace s_i=\bar{1},1\right\rbrace$ is known to be given by the principles of equilibrium statistical mechanics.
\begin{subequations}\label{PnEn}
   \begin{align}
	P_n&=\frac{e^{-E_n/k_BT}}{Z}                 \label{Pn} \\
    \intertext{where the energy $E_n$ of configuration $n$ is given by}
	E_n&\approx -\Sigma_{i,j} w_{ij} s_i s_j      \label{En}
    \end{align}
\end{subequations}
\par
\paragraph{} Ising models are closely related to Boltzmann machines (\cite{Smolensky, Hinton1985, BengioAI, Hinton2006NC, Hinton2006Sc, BengioGreedy}) whose
probabilities described by Eq.\ref{Pn} seek configuration with low $E_n$. For example, with three transynapses connected through
$w_{ij}=w_{ji}>0$, $E_n$ is minimized for configurations $(111)$ and $(\bar{1}\bar{1}\bar{1})$ with equal ${s_i}$. This is the
ferromagnetic (FM) Ising model. But if $w_{ij}=w_{ji}<0$, $E_n$ would be a minimum if all ${s_i}$ had opposite signs. Since this is
impossible with three transynapses, the energy is lowest for all six configurations that have one `frustrated' pair
\cite{BookSG,TheoryRM}:
\begin{equation}
A:\bar{1}11,1\bar{1}\bar{1} \hspace{0.5cm}  B:\bar{1}1\bar{1},1\bar{1}1 \hspace{0.5cm} C:\bar{1}\bar{1}1,11\bar{1} \label{ConSpa}
\end{equation}
\par
\begin{figure}[]
  \centering
  \includegraphics[width=14cm, height=7cm]{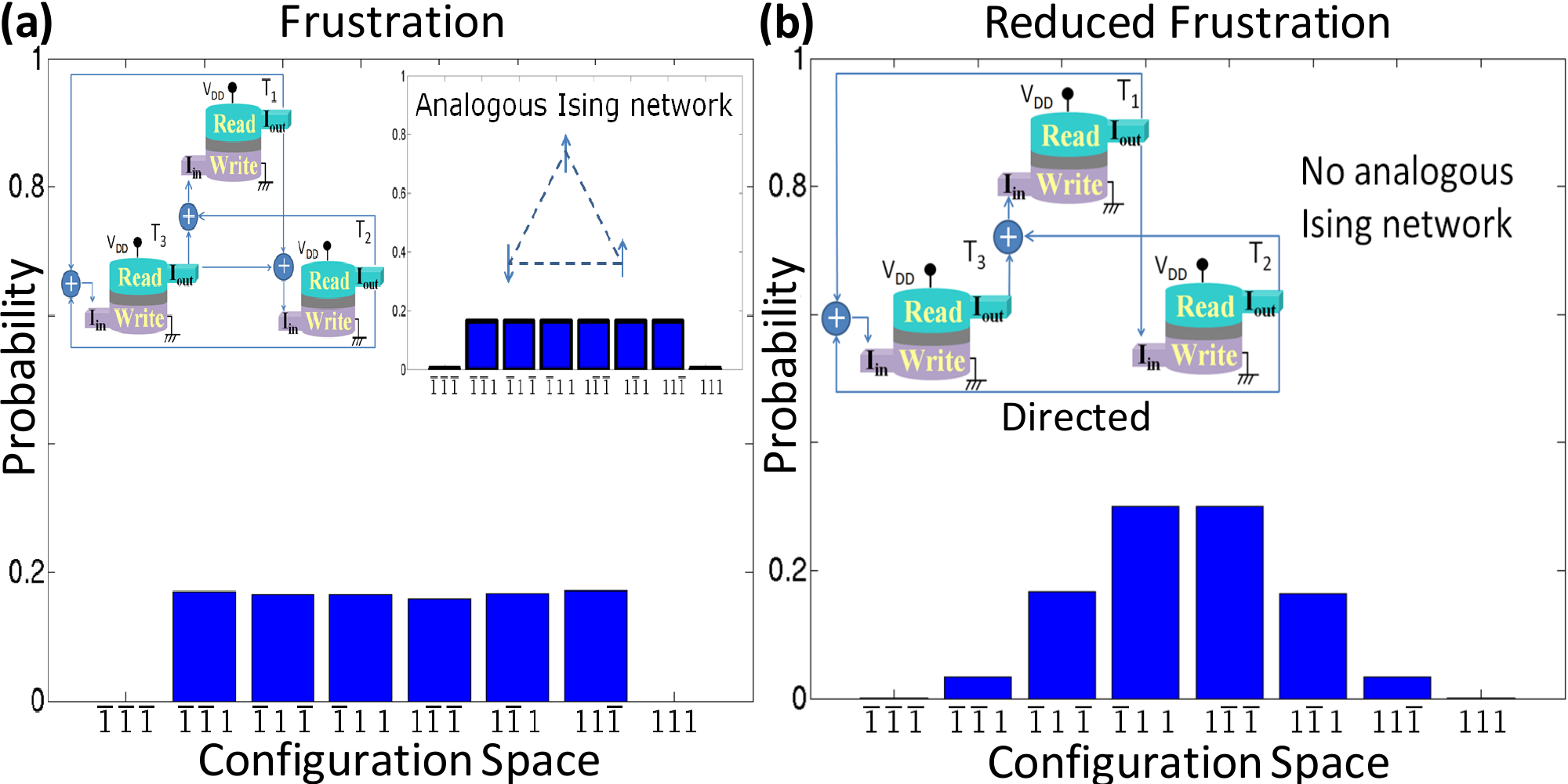}\\
  \caption{\textbf{(a)} Three Transynapses are initialized and then left to relax while interacting in a pairwise manner. The
  strength of interactions depends on voltage $V_{DD}$. The polarity of $V_{DD}$ for each transynapse is such that it favors the
  next transynapse to have an opposite state to its own as in anti-ferromagnetic (AF) ordering. Statistical information is then
  gathered from Monte-Carlo runs. There are a total of $2^3$ configurations possible with their probabilities shown in Fig.\ref{F3}
  a. This is reminiscent of frustration in spin glasses \cite{BookSG,TheoryRM} also observed in Ising model as shown in the inset. Such
  bidirectional connections can be used for building Boltzmann machines\cite{Smolensky,Hinton1985} closely related to Ising models.
  More on this in Fig.\ref{F4}. \textbf{(b)} Changing one of the connections in part (a) to be directed as opposed to bi-
  directional lowers the probability of occurrence of some states in the final configuration resulting in reduced frustration. This
  is fundamentally not possible by inherently symmetric Hamiltonian based systems such as Ising model. Such directed
  connections can be used to represent causal influences in Bayesian networks \cite{PearlBook, PearlCausality}. More on this in
  Figs.\ref{F5},\ref{F6}.}
  \label{F3}
\end{figure}

\paragraph{} The numerical simulation of the 3-transynapse network shows (Fig.\ref{F3}a) this expected behavior
with equal probabilities for configurations A,B,C, and reduced probabilities for the two remaining configurations $(111)$ and $
(\bar{1}\bar{1}\bar{1})$ for which all three pairs are frustrated. Situation is different when one of the bonds is directed as in
Fig.\ref{F3}b. Not surprisingly, the probability is highest for the configuration having $T_2$ and $T_3$ as the frustrated pair
(configuration A in Eq.\ref{ConSpa}). Less obviously, configuration B with $T_1$, $T_3$ as the frustrated pair has a higher
probability than configuration C with $T_1$, $T_2$ as the frustrated pair. This is because $T_2$ only has one bond (from $T_1$)
dictating its state (no conflict) but $T_3$ has two bonds (from $T_1$ and $T_2$) dictating its state which can be at odds with each
other. Such configuration of bonds and the resulting configuration space probabilities have no Ising analog. \par
\paragraph{} Note that our numerical results are all obtained directly by simulating a set of coupled LLG
equations, one for each of the six magnets, two per transynapse. The time evolution of each magnet in each device is a function of
its instantaneous state  $\vec{M}_i(t)$, internal, external and thermally fluctuating fields (determined by temperature $T$), plus
the spin torque $\vec{\tau}_{ij}$ it receives from other devices:
\begin{equation}
d\vec{M}_i/dt=f( \vec{M_i},\vec{H_i}_{int},\vec{H_i}_{ext},\vec{H_i}_{flc}(T),\Sigma_j\vec{\tau}_{ij}(\vec{M}_i) ) \label{LLG}
\end{equation}
Bi-directional interactions have both $\vec{\tau}_{ij}$ and $\vec{\tau}_{ji}$ but directional interactions have either $\vec{\tau}_{ij}$ or $\vec{\tau}_{ji}$. Supplementary section 1 provides more detail. \par

\pagebreak

\section{Implementing Belief Networks}
\subsection{Boltzmann machine}
\begin{figure}[]
  \centering
  \includegraphics[width=9cm, height=7.5cm]{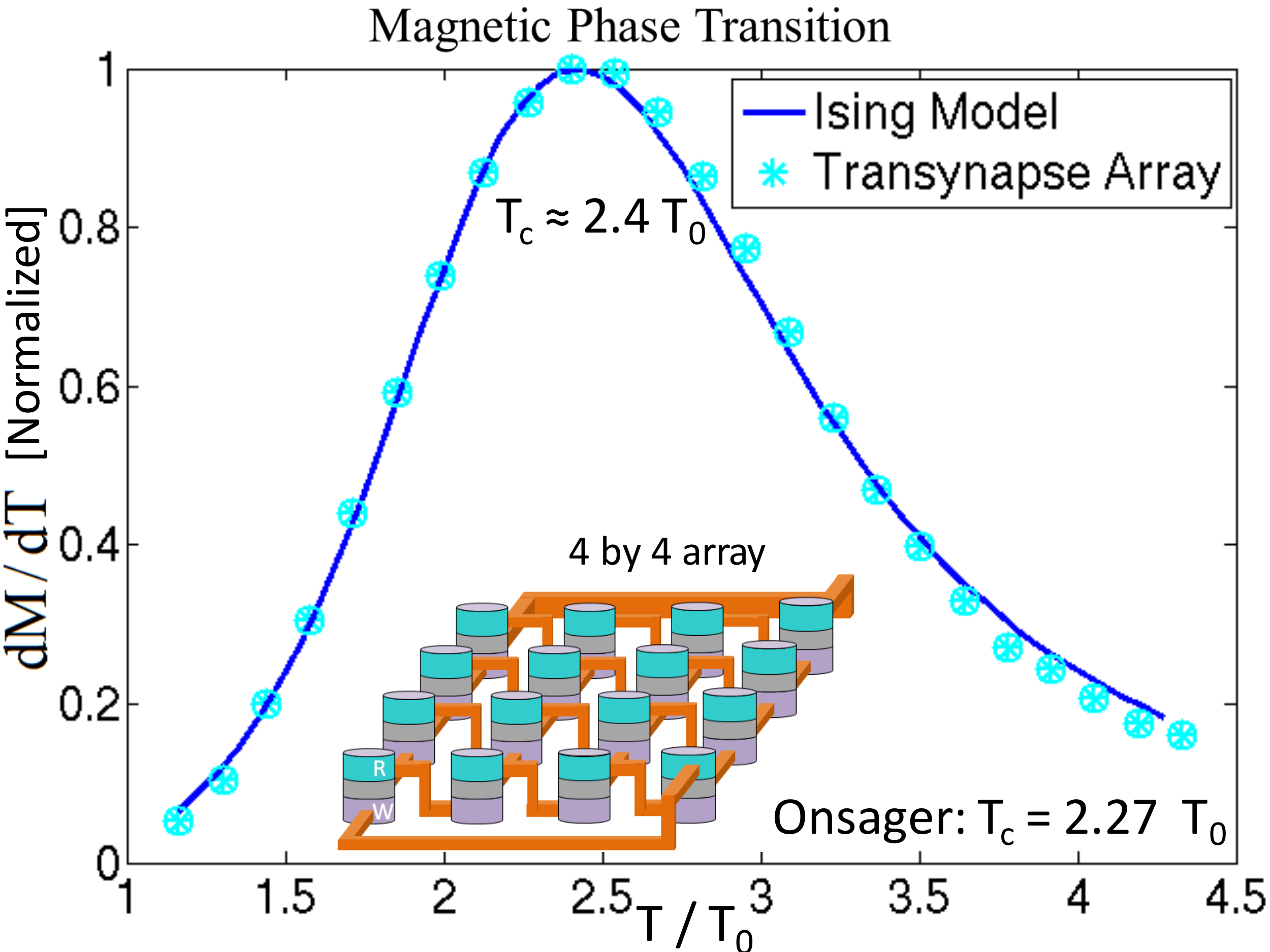}\\
  \caption{A 4 by 4 array of transynapses with nearest neighbor connections (The inset is intended to illustrate these aspects and not the directionality of connections). Same $V_{DD}$ ($V_{DD}<0$) is applied to all transynapses making the interactions favor all devices in the same state similar to ferro-magentic ordering (FM). At each temperature $T$ (scaled by $T_0 \equiv J/k_B$, where $J$ is the coupling strength. See also supplementary sections 3 and 4), the circuit is initialized and left
  to interact while the network decides on a final state out of $2^{16}$ possible states. After each trial, the magnetization of the array is
  obtained by summing over all transynapse states leading to an average magnetization of the array based on the total number of Monte Carlo
  runs. From this data, differential of magnetization with respect to temperature can be obtained as shown.
  This is reminiscent of magnetic phase transition \cite{Landau} exhibiting a Curie temperature in Ising model depicted by the solid line. In
  deep belief networks \cite{Hinton2006NC,Hinton2006Sc, Hinton2002PoE,CDBN}, the closely related restricted Boltzmann machines \cite{Smolensky,Hinton1985} trained by breakthrough algorithms that determine the interactions are used to solve search, detection and inference problems.
  \cite{BengioGreedy, Hinton2002PoE,BengioReview}}
  \label{F4}
\end{figure}

\paragraph{} The connection between Ising model of statistical mechanics \cite{Cipra} and hard combinatorial optimization problems of mathematics has been known for decades.\cite{Kirkpatrick} Boltzmann machines \cite{Smolensky,Hinton1985} and subsequently their restricted version for deep belief networks are Ising models in which the weights of interactions are learned and adjusted with breakthrough algorithms.\cite{Hinton2006NC, Hinton2006Sc, BengioGreedy} There is also widespread activity and innovation on the connection of inference, commonly used in belief networks, and phase transitions in statistical physics (see e.g. Ref.\cite{StatInfe} for a thorough review). Figure \ref{F4} shows how networks described in this paper (Fig.\ref{F1}) can mimic magnetic phase transition which is also a well known result of the Ising model. The caption provides more detail for the particular procedure used for obtaining this. Phase transition is evident as the rate of change of magnetization with respect to temperature exhibits a maximum followed by a decrease. This transition is not sharp because of the small lattice sizes used here (see Ref.\cite{Landau} for a more in-depth discussion). Solid line shows the analogous Ising model result with the same lattice size (4 by 4 array) using equilibrium laws of statistical mechanics. The peak exhibited is reminiscent of the Curie temperature of magnetic phase transition (supplementary section 3). Indeed, the effective Curie temperature observed in these networks depends linearly on the strength of device to device communication set by $V_{DD}$ (Fig.\ref{F2}). (Supplementary section 4 provides spontaneous magnetization curves with interactions of various strength). This is in agreement with Onsager's \cite{Onsager} results for a two dimensional array of ferromagnetic atoms for which $T_C$ is proportional to $J$-coupling strength. We take these as an indication that stochastic networks of transynapses could be used to construct (restricted) Boltzmann machines for deep belief networks where weights can be adjusted by the bipolar voltages applied to transypases or by load resistances at the output of transynapses. (Fig.\ref{F2})\par

\pagebreak
\subsection{Bayesian network}
\begin{figure*}[]
 \centering
  \includegraphics[width=7cm, height=8.75cm]{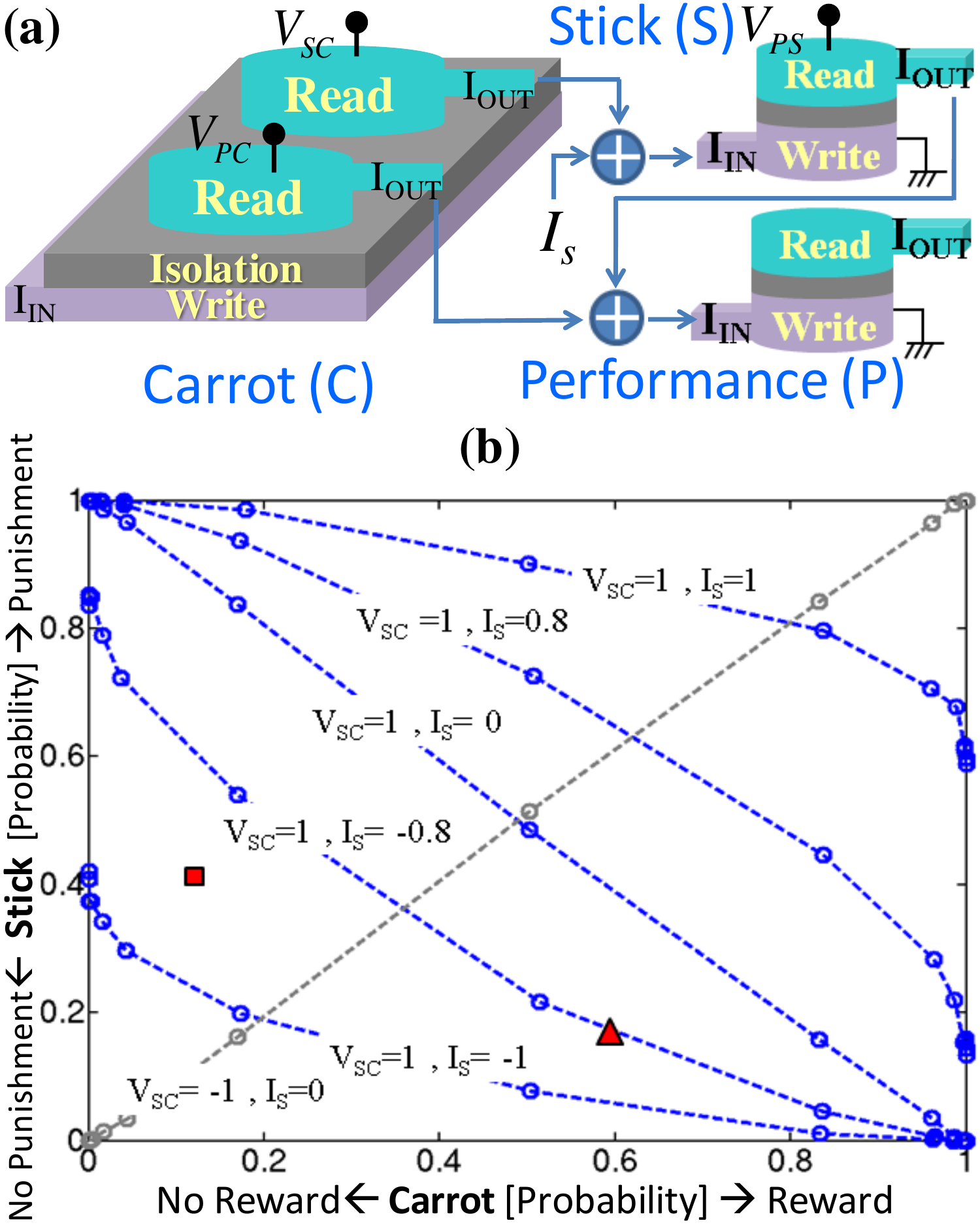}\\
  \caption{\textbf{(a)} Directed circuits of Transynapses (Fig.\ref{F1}) can represent
  Bayesian networks in which directionality can represent causality . \cite{PearlBook, PearlCausality}
  Unlike Hamiltonian systems (e.g. Ising model), the interactions are not symmetric. Here,
  carrot influences both the state of the stick and the performance while stick also affects
  performance. \textbf{(b)} Direct simulation of Fig.\ref{F5}a. Figure shows a probabilistic gate in which
  diagonal lines represent perfect correlation of stick with carrot (probabilistic COPY) using $V_{SC}=-1$ and
  perfect anti-correlation (probabilistic NOT) using $V_{SC}=+1$ (voltages and currents
 are normalized by the magnitude required for deterministic switching). When $I_S\neq 0$, statistical correlation varies
  e.g. the stick can be in the no-punishment mode irrespective of the carrot.}
   \label{F5}
\end{figure*}

\paragraph{} Symmetric interactions are inherent to Hamiltonian based systems as in Ising model and Boltzmann machines. On the other hand, directed interactions have their own prominence in Bayesian networks \cite{PearlBook, PearlCausality}. Figure \ref{F5}a shows a 3-transynapse network, with each transynapse representing one of three variables which we could call carrot, stick and performance. These variables can be in one of two possible states

\begin{center}
\begin{tabular}{l | c | c | c}
          & Carrot (C)    &  Stick  (S)    & Performance (P)  \\
\hline
1         & Reward    &  Punishment    & Better       \\
$\bar{1}$ & No Reward &  No Punishment & Worse        \\
\end{tabular}
\end{center}
with distinct probabilities. The transynapse network is interconnected to reflect the causal
interconnections among the three variables. The carrot affects both the state of stick and the state
of performance through the voltages $V_{SC}$ and $V_{PC}$ which determine the weights $w_{SC}$ and
$w_{PC}$. The only other causal effect is that of the stick on the performance which is reflected in
the voltage $V_{PS}$ and the resulting weight $w_{PS}$. \par
\paragraph{} A direct simulation of this 3-transynapse network using coupled LLG equations (Eq.
\ref{LLG}) yields the plot shown in Fig.\ref{F5}b. With $V_{SC}=-1$ (voltages and currents
are normalized by the magnitude required for deterministic switching), we get the diagonal lines
reflecting perfect correlation of the stick with the carrot, while with $V_{SC}=+1$, we get the other
diagonal line reflecting perfect anti-correlation. We could view these respectively as a COPY gate
and a NOT gate with probabilistic inputs and outputs. The other curves shown in Fig.\ref{F5}b
correspond to $I_S\neq 0$ reflecting a situation where the stick state is not entirely controlled by
the carrot, but has a probability of no punishment irrespective of the carrot.\par

\paragraph{} The network can naturally generate probabilities of various variables. Consider e.g. the
triangle (scenario A) in Fig.\ref{F5}b where carrot has 0.6 probability of reward (scenario B is the
square). Instead of performing the necessary algebra of $p(S=1) = \Sigma_{C\in\left\lbrace 1,\bar{1}
\right\rbrace} p(S=1|C)$ to obtain the probability of stick being in the punishment mode, the
transynapse network takes in $I_C=-0.02$ , $ V_{SC} = 1$, $I_S=0.9$ and produces the directly
observable probability of stick being in punishment mode. This generalizes to more variables and an
example for three is discussed next.\par
\begin{figure*}[]
 \centering
  \includegraphics[width=8.4cm, height=7cm]{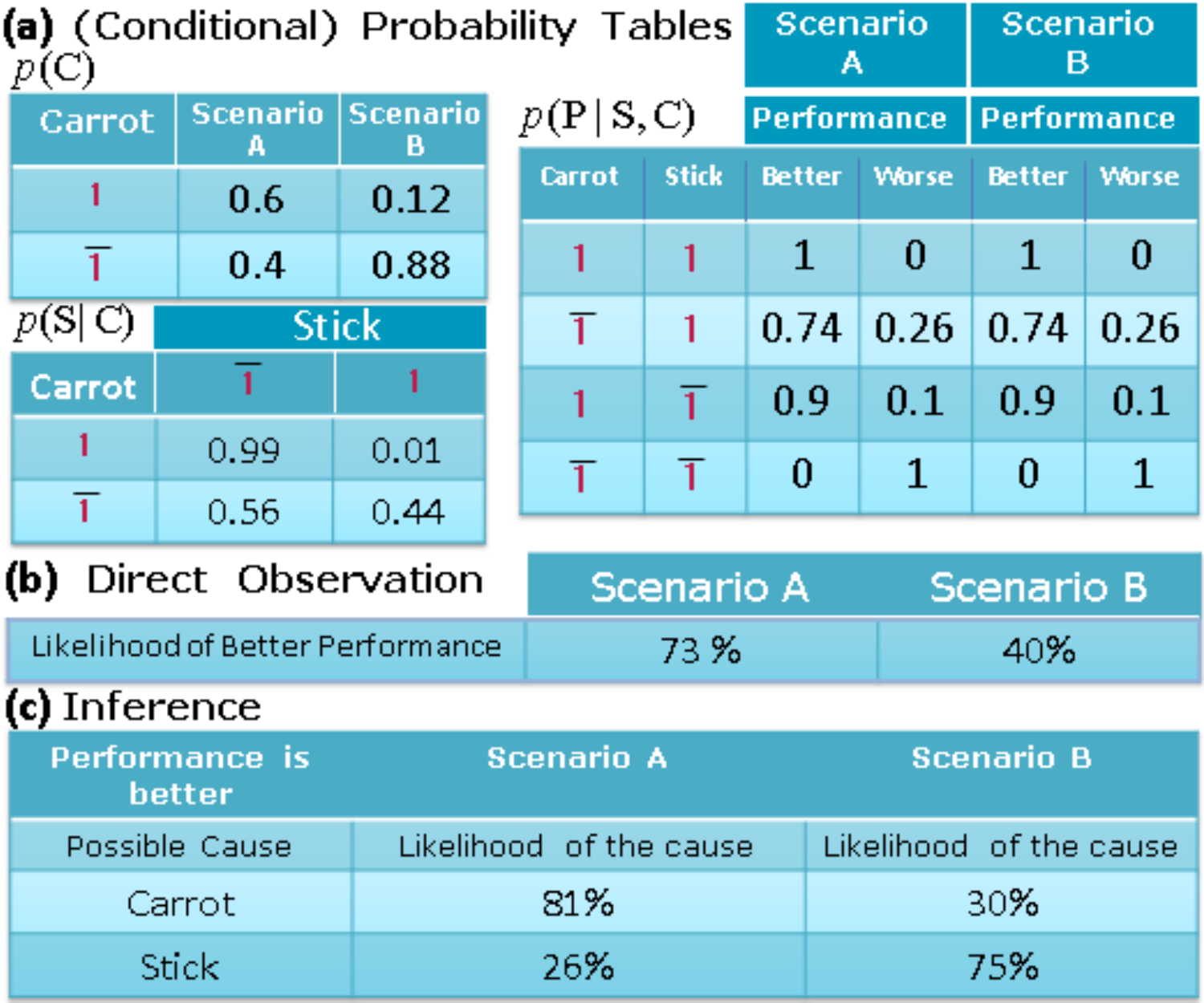}\\
  \caption{\textbf{(a)} (Conditional) probability tables: Two scenarios (A:triangle and B:square in Fig.\ref{F5}b) are
  considered for the state of carrot. Such scenarios are typically provided by the problem statement
  which determines the voltages and currents applied to transynapses based on their transfer function
  (Fig.\ref{F2}b). They in turn ensure that the network generates the probability values as shown.
  $V_{SC}=1, V_{PC}=-0.4,  V_{PS}=-0.5, I_C=-0.02, I_S=0.9, I_P=-0.2$ are used for scenario A . Same
  values are used for scenario B except that $I_C=0.1$ (voltages and currents are normalized by the
  magnitude required for deterministic switching). \textbf{(b)} Likelihood of better performance can be
  directly observed  without using Fig.\ref{F6}a and carrying out the algebra for $p(P=1|S,C)$.
  \textbf{(c)} Inference can be addressed by such networks. For example, the likelihood that reward has
  caused better performance is the mean value of carrot in the reward state for cases that have better performance.}
   \label{F6}
\end{figure*}

\paragraph{} Figure \ref{F6}a,b shows how the network in Fig.\ref{F5}a can be used in predictive mode based on known casual
connections \cite{PearlBook, PearlCausality} among different variables which determine the electrical signals $V_{ij}$ and $I_i$ (explicitly provided in the caption). These in turn can provide the values in the (conditional) probability tables of Fig.\ref{F6}a. For example, the element indexed by (1,$\bar{1}$) in the $p(S|C)$ table is the mean value of the state of stick in the $\bar{1}$ mode when $C=1$. (This can also be obtained independently by dictating the carrot is in the reward ($C=1$) state e.g. by providing a strong bias ($I_C$) and finding the mean value for the state of stick due to $V_{SC}$) While the likelihood of better performance can be found from tables of Fig.\ref{F6}a by calculating $p(P=1|,S,C)$, this is directly observable from the mean value of the state of `performance' which is naturally generated by the network as provided in Fig. \ref{F6}b. Alternatively, the network can address inference problems. Suppose performance is better, is it due to carrot or stick or both?  For instance, the likelihood that performance is better because of reward is essentially $p(C=1|P=1)$. This can be obtained by the algebra, $\Sigma_S \hspace{0.2cm} p(P=1,S,C=1) \hspace{0.1cm} / \hspace{0.1cm} \Sigma_{C,S} \hspace{0.2cm} p(P=1,S,C)$, or directly observed by taking the mean value of the state of carrot in the reward mode when performance is better. The resulting values are provided in Fig.\ref{F6}c.\par

\section{Concluding Remarks}
\paragraph{} Probabilistic computing is a thriving field of computer science and mathematics that deals with extracting knowledge from available data to guide decisive action. The work in this
area is largely based on deterministic hardware and major advances can be expected if one
could build probabilistic hardware to simulate probabilistic logic. In this paper we define
a building block for such stochastic networks, which we call a transynapse combining the
transistor-like properties of gain and isolation with synaptic properties. We present a possible
implementation based on the established physics of nano magnets. Using experimentally
benchmarked models for the transynapse we present examples illustrating the implementation
of both Boltzmann machines and Bayesian networks. More realistic examples will be addressed in future publications \cite{Brian}.\par

\section{Acknowledgment}
VD was supported by the Center for Science of Information (CSoI), an NSF
Science and Technology Center, under Grant agreement No. CCF-0939370.


\pagebreak
\section{Supplementary Information}
\subsection{Methods and verification}
\paragraph{} This section outlines the methodology and its implementation underlying the simulations that have been carried out. It includes the steps taken to verify the model against well known principles or experimental data. \\

The time evolution and final state of each nano-magnet $\vec{M}=(M_s\Omega)\hat{m}$ is represented and simulated by the Landau-Lifshitz-Gilbert (LLG) equation:
\begin{equation}
\label{LLG}
\frac{d\hat{m}}{dt}=-|\gamma|\hat{m}\times\vec{H}
+ \alpha\hat{m}\times\frac{d\hat{m}}{dt}
- \frac{1}{qN_s}\hat{m}_n\times\left(\hat{m}\times\vec{I}_{s}\right)
\end{equation}

where $q$ is the charge of electron, $\gamma$ is the gyromagnetic ratio,
$\alpha$ is the Gilbert damping coefficient and $N_s\equiv M_s\Omega/\mu_B$ ($M_s$: saturation magnetization, $\Omega$: volume) is the net number of Bohr magnetons comprising the nanomagnet and $\vec{I}_s$ is the spin current entering the magnet.\\

This equation is transformed to its standard mathematical form (see e.g. Ref.\cite{ASL}) and is solved numerically using a second order Runge-Kutta method (a.k.a Heun's method) in MATLAB. This methodology essentially applies the Stratonovich stochastic calculus to the stochastic integration during time dependent simulations involving thermal fluctuations. The inclusion of thermal fluctuations in LLG and its implementation has been verified against equilibrium laws of statistical mechanics in Ref.\cite{ASL}.\\

$\vec{I}_s$, the spin current entering the magnet, can have components both due to the Slonczweski torque as well as the field-like torque. The inclusion of spin transfer torque in LLG (last term)  and its implementation has been verified against experimental data in Ref.\cite{BehtashAPL}. In this manuscript, $\vec{I}_s=\beta I \hat{z}$ is generated by the spin Hall effect as outlined in \cite{DattaSS} where $I$ is the charge current entering the $W$ unit generated by the READ stage of the previous device (see Figures 1-3). \cite{DattaSS} this is essentially how Transynapses communicated with each other whereby one drives the next; the dynamics of each one being governed by the coupled LLG equations that describe the dynamics of READ and WRITE magnets.\\

The magnetic field, $\vec{H}$, represents both internal and external fields:
\begin{eqnarray}
\label{H-Fields}\nonumber
\vec{H} &=& \vec{H}_{int} + \vec{H}_{ext}        \\  \nonumber
\vec{H}_{int} &=& \vec{H}_u + \vec{H}_{demag} \\  \nonumber
\vec{H}_{ext} &=& \vec{H}_{fl} + \vec{H}_{couple}    \nonumber
\end{eqnarray}
where $\vec{H}_u = H_Km_z\hat{z}$ is the uniaxial anisotropy field with $z$ as the easy axis and  $\vec{H}_{demag}= - H_dm_y\hat{y}$ is the demagnetizing field with $y$ as the out of plane hard axis for in-plane magnets. $H_d$ is zero for perpendicular anisotropy magnets.\\
The thermal fluctuating field, $\vec{H}_{fl}$, has the following statistical properties:
\begin{eqnarray}
\label{First Moment}
\left<H^i_{fl}\left(t\right)\right>&=&0\\  \nonumber
\label{Second Moment}
\left<H^i_{fl}\left(t\right)H^j_{fl}\left(t^{\prime}\right)\right>&=&\delta_{ij}\delta\left(t-t^{\prime}\right)\sigma^2\\ \nonumber
\label{Variance}
\sigma^2 &=& \frac{\alpha}{1+\alpha^2}\frac{2k_BT}{|\gamma| M_s\Omega} \nonumber
\label{Stat Prop}
\end{eqnarray}
where $\delta(t)$ is the Dirac delta function, $\delta_{ij}$ is the Kronecker delta,
and indices $i$ and $j$ are labels for the field's vector components. $T$ is temperature and $k_B$ is the Boltzmann constant.\\
\begin{figure*}[]
  \centering
   \includegraphics[width=14cm, height=12cm]{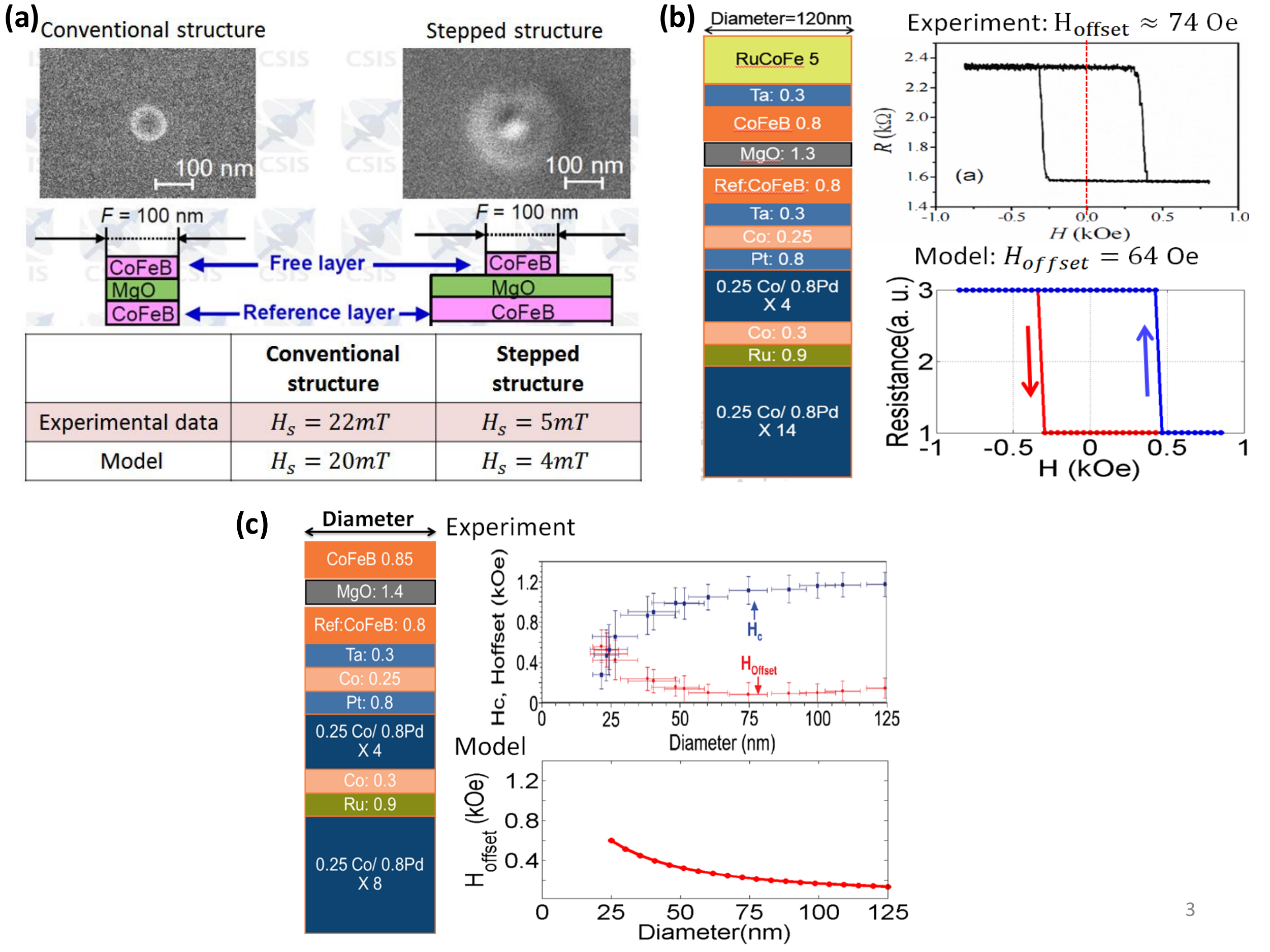}\\
  \caption{\textbf{(a)} Model is validated against reference \cite{Tohoku} for the effect of the stray fields in shifting the R-H hysteresis
  loops for two different structures as described in the reference. \textbf{(b)} Model is validated against reference \cite{IBMShift} for the
  effect of the stray fields in shifting the R-H hysteresis loops in MTJ structure as illustrated layer by layer on the left. \textbf{(c)}   	
  Model is validated against reference \cite{IBMScaling} for the scaling of coupling fields as the diameter of magnetic tunnel junctions are
  scaled.}
  \label{FS1}
\end{figure*}

\indent The coupling field, $\vec{H}_{couple}$, accounts for the magnetic interaction of the READ (R) and WRITE (W) magnets within each device as introduced and described in \cite{DattaSS} and illustrated in Fig.2a of the main manuscript. (There is no magnetic coupling envisioned between various devices here. Device to device communication happens via charge currents as described earlier.) This reference describes functionality of the spin switch and the governing equations and presents the coupled LLG equations describing the time dynamics of R and W magnets (see its supplementary section). The modeling of magnetic coupling between READ and WRITE magnets for in-plane magnetic materials has been described in detail in \cite{VinhDis} and verified against experimental data. Here, we review a brief description of how this coupling is calculated for perpendicular magnetic materials along with the validation of its implementation against experimental data from various experiments in Fig.\ref{FS1}. We follow the methodology described in \cite{Newell}. Within each device, W and R magnets exert a magnetic field on the other. For example,  the field exerted on the READ magnet from the WRITE magnet is $\vec{H}_{RW}= [D]M_{s,W}\vec{m}_W$ where [D] is a 3 by 3 tensor describing the effect of each elemental volume of W magnet on each elemental volume of the R magnet integrated over the volume of both
\begin{equation}
\label{D}\nonumber
[D]_{ij} = \frac{1}{4\pi} \int_{\Omega_{W}}\nabla_i \left( \int_{\Omega_R} \nabla_j \left(\frac{1}{|\vec{r}-\vec{r}'|} \right) dr' \right) dr
\hspace{1cm} i,j=x,y,z
\end{equation}
To validate the approach and its implementation, we made use of the available experimental data for the coupling fields that have been measured in magnetic tunnel junctions. Figure $S1$ shows the comparisons between the calculated values from the model and the data from various experiments. These comparisons show that the model is generally in good agreement with experimental demonstrations.

\subsection{Model specifications and parameters}
The spin switch consists of various layers as discussed in Ref.\cite{DattaSS}. The free coupled magnetic layers are nominally identical. Their dimensions are $100*100*2 nm^2$ with $M_s=1000$ emu/cc and $H_K=200$ Oe for the $E_b\approx48kT$ magnet and $H_K=50$ Oe $E_b\approx12kT$ magnet in Fig.2. Both have damping parameter of $\alpha=0.01$. Both perpendicular anisotropy  magnets and in-plane magnets can produce the stochastic sigmoid function like Fig.2. Figures 3,5 and 6 use lower energy barrier magnets which due to lower volume (100*50*2 $nm^3$) and anisotropy field ($H_K=25$ Oe). The separation between these coupled magnets is $7$ nm. Figure 4 which shows the magnetic phase transition uses magnets with very small barriers. This is to be more consistent with Ising spins which actually do not have any intrinsic energy barriers. The free layer specifications used in the transynapse network of figure 4 are 50*50*2 $nm3$ and $H_K=20$ Oe. The ``$J$-coupling'' for the transynapse network was adjusted by $V_{DD}$ to achieve a current that is twice the switching current of the each transynapse. This is the maximum magnitude of the current $I_{cpl}$ that each transynapse sends to other ones. The connection between the $J$-coupling of the Ising model (discussed next) and the transynapse network was made by setting the $J$ of transynapse network to to $I_{cpl}/4$.
\subsection{Ising Model}
The Ising model is a well-known established mathematical model of ferromagnetism in statistical mechanics. Where applicable, in the main manuscript, it has been used to draw comparison between networks of Ising spins and magnetic networks presented in this manuscript. For that purpose, a MATLAB script was written based on the governing equations outlined below. The standard form of this model describes the interaction energy or the Hamiltonian of a network of Ising spins which can be under the influence of an external magnetic field. For a network in state $S$, the many-body Hamiltonian of size $2^n$ by $2^n$ where $n$ is the number of Ising spins can be written as
\begin{equation}
\label{HS}\nonumber
H_S = - \Sigma_{ij} J_{ij}S_iS_j - \mu \Sigma_ih_iS_i \hspace{1cm} S_{i,j}\in \left\lbrace+1,-1\right\rbrace
\end{equation}
%
Where $J_{ij}$ describes the interaction energy between two nearest neighbor Ising spins $S_i$ and $S_j$. Magnetic moment of each spin is denoted by $\mu$ and could be under the influence of an external magnetic field $h_i$. Depending on the values assumed by the Ising spins, the network can have various possible configurations (states).  The probability of each state $S$ at equilibrium in the configuration space is given by
\begin{equation}
\label{PS}\nonumber
P_S = \frac{e^{-\beta H_S}}{Z} \hspace{1cm} \beta \equiv (k_BT)^{-1}
\end{equation}
%
$k_B$ is the Boltzmann constant and T is the ambient temperature. $Z$ is the well-known partition function of equilibrium statistical mechanics $Z = \Sigma_S e^{-\beta H_S}$. Note that $\Sigma_sP_s=1$ always.\\

\begin{figure*}[]
  \centering
   \includegraphics[width=10cm, height=8cm]{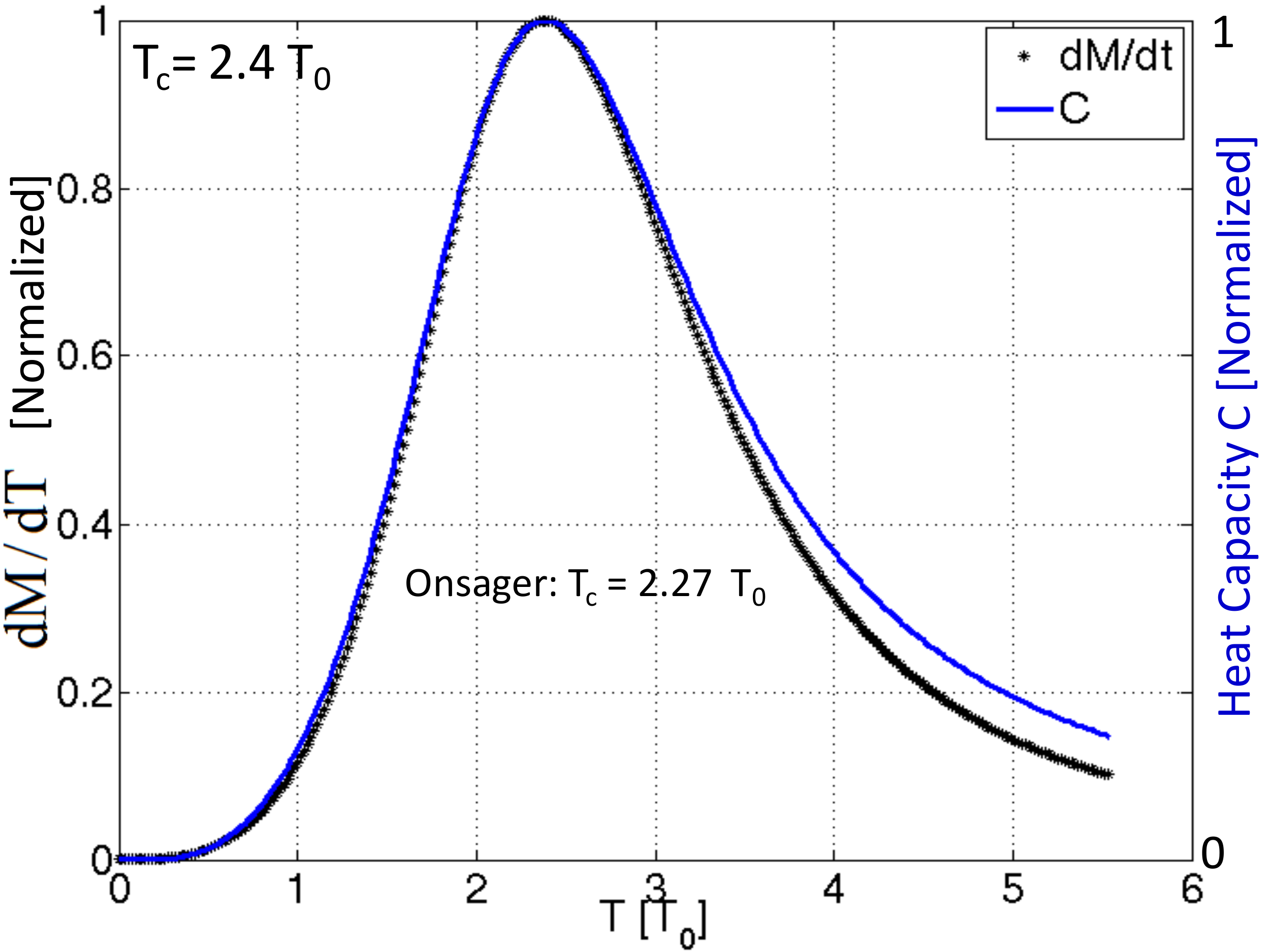}\\
  \caption{Heat capacity and rate of change of magnetization (phase) with respect to temperature (order parameter) plotted as a functional of temperature. Phase transition and Curie temperature are evident from both curves.}
  \label{FS2}
\end{figure*}

Energy and magnetization of such a system are observable and can be readily found by taking the their expectation value
\begin{equation}
\label{E}\nonumber
E = \Sigma_S H_S P_S
\end{equation}
%
\begin{equation}
\label{M}\nonumber
M = \Sigma_S M_S P_S
\end{equation}
%
Here, $M_s$ is the normalized net magnetization of each state of the many-body system found by summing over all Ising spin values of each state.
For magnetic phase transition, we look at the heat capacity as a function of temperature. Heat capacity, $C$, is a measure of how much heat is added to a system for a given temperature change
\begin{equation}
\label{M}\nonumber
C = \frac{dE}{dT}
\end{equation}
Curie temperature, $T_C$ is the critical point at which the heat capacity ($C$) peaks while the material's phase (magnetization here) exhibits an inflection point; hence the phase change. It is easily recognizable in the $C$ versus $T$ plot (supplementary Fig.\ref{FS2}). But it also is recognizable from the rate of change of materials' phase (magnetization) with respect to the order parameter (temperature here). This is discussed more in detail in the next section.

\begin{figure*}[]
  \centering
   \includegraphics[width=10cm, height=8cm]{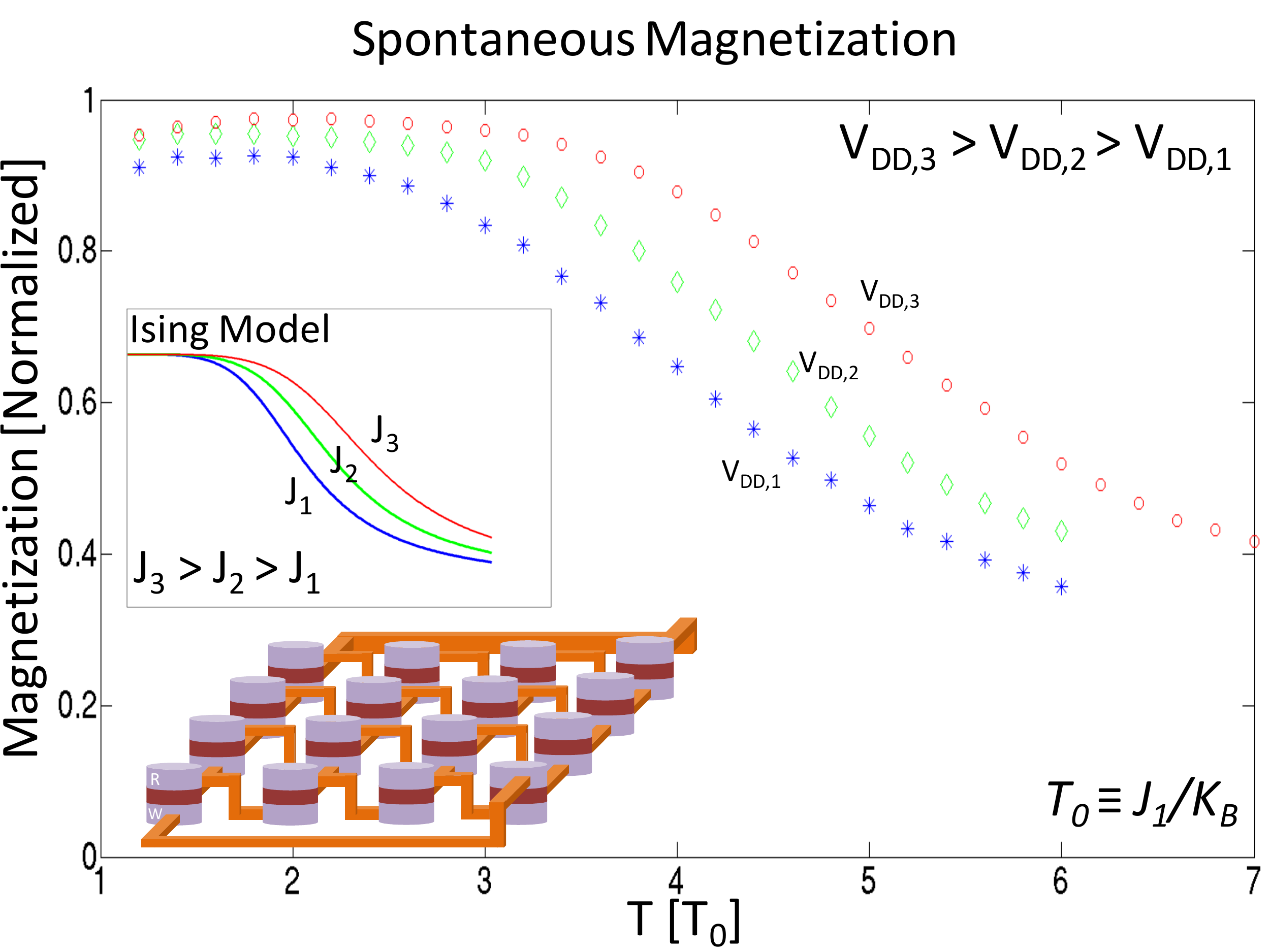}\\
  \caption{Effective magnetization in a network of Transynapses. Changing $V_{DD}$ applied to each Transynapse can control the strength of
  the interaction between them somewhat like the J-coupling of Ising spins. Similar to the work of Ansager in describing $T_c$, spontaneous
  magnetization curves shift linearly as a function of the strength of the interactions. Higher $V_{DD}$ makes the network to retain its
  magnetization at higher temepratures effectively raising the effective $T_c$ in a linear fashion.}
  \label{FS3}
\end{figure*}
\subsection{$V_{DD}$ tuning of spontaneous magnetization and control of effective Curie temperature $T_c$}
Supplementary Fig.\ref{FS3} shows spontaneous magnetization as a function of temperature for a transynapse network and Ising model (both are 4 by 4 arrays). The inflection point where the magnetization changes its curvature marks the Curie temperature point $T_C$ which is the point at which the rate of change of magnetization peaks (supplementary Fig.\ref{FS2}). What Fig.3 shows is that for a linear increase in $V_{DD}$, spontaneous magnetization curves are shifted linearly. This is in agreement with Onsager's derivation, $T_C=2.27 J/k_B$, essentially explaining the linear dependence of $T_C$ on the coupling strength $J$ between the Ising spins. To clarify this further, the inset  shows an analogous plot based on the Ising model illustrating the linear shift of spontaneous magnetization curves. Note also that for Figure 4 of the main paper, the data for magnetization has been smoothed out using the MATLAB function ``smooth'' before taking the derivative.

\pagebreak

\end{document}